\documentclass[TOTEM]{cernphprep}

\usepackage{fancyhdr, graphicx}
\usepackage{float}
\usepackage{color}
\usepackage{amsmath,amssymb,wasysym}
\usepackage{multirow}
\usepackage{epstopdf}
\usepackage{lineno}
\usepackage[numbers,sort&compress]{natbib}
\usepackage{url}

\DeclareMathAlphabet{\mathcal}{OMS}{cmsy}{m}{n}

\def\un#1{\,{\rm #1}}

\setbox123\hbox{\small$0$}
\def\S{\hbox to\wd123{\hss}}
\setbox124\hbox{\small$_{0}$}
\def\s{\hbox to\wd124{\hss}}

\def\etal{et al.}

\def\Instline#1#2{%
	\expandafter\write1{\string\newlabel{#1}{{#1}{}}}%
	\hbox to\hsize{\strut\hss$^{#1}$#2\hss}
}

\begin{document}

\begin{titlepage}

\renewcommand{\EXPLOGO}{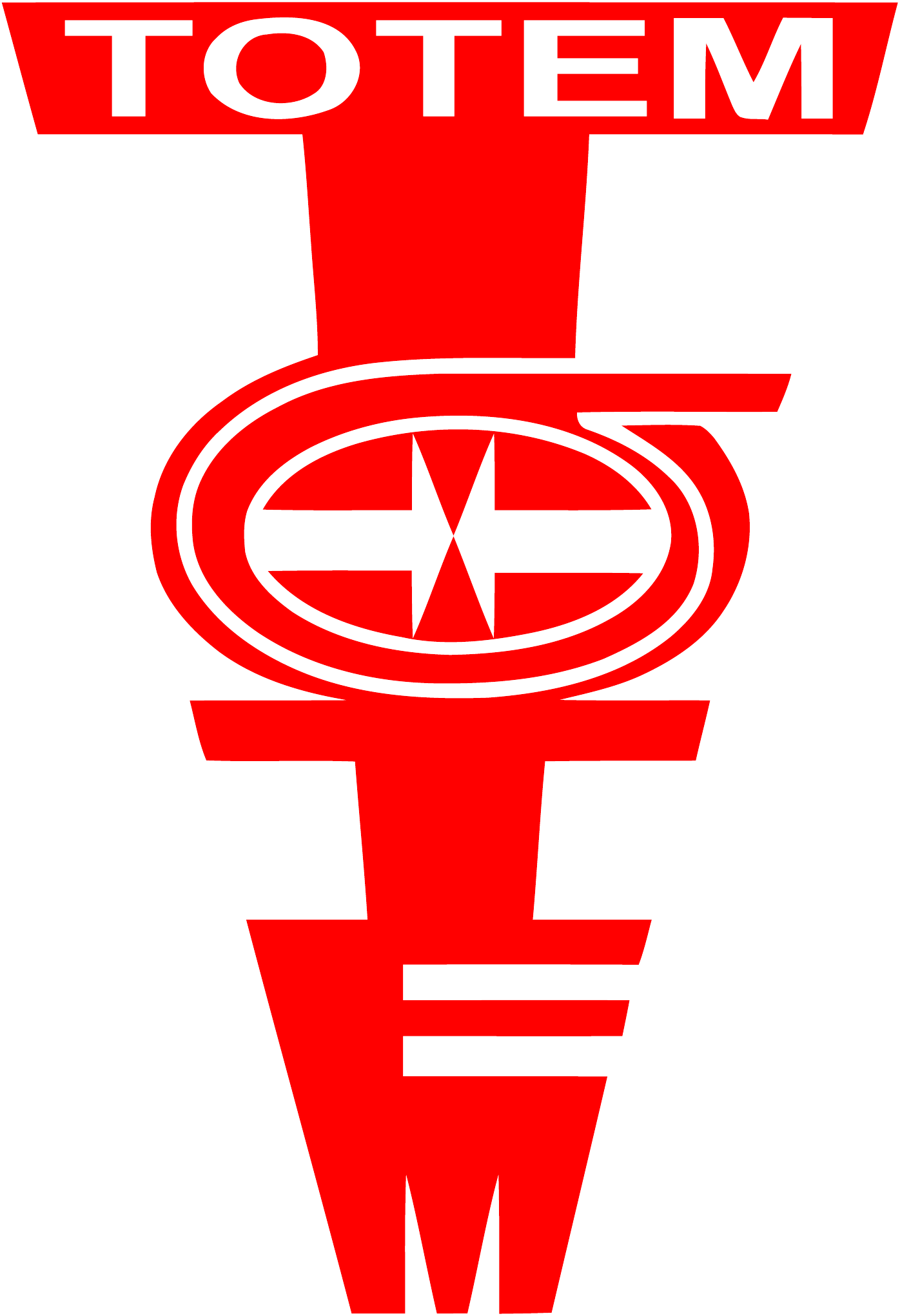}

\PHnumber{CERN-EP-2018-341}
\PHdate{\today}

\EXPnumber{TOTEM-2018-002}
\EXPdate{\today}

\title{Elastic differential cross-section ${\rm d}\sigma/{\rm d}t$ at $\sqrt{s}=$2.76 TeV and implications on the existence of a colourless 3-gluon bound state}
\ShortTitle{Proton-proton differential cross-section at $\sqrt s = 2.76\un{TeV}$}

\Collaboration{The TOTEM Collaboration}
\ShortAuthor{The TOTEM Collaboration (G.~Antchev \emph{\etal})}

\def\DeclareAuthors{%
	\AddAuthor{G.~Antchev}{}{}{a}%
	\AddAuthor{P.~Aspell}{9}{}{}%
	\AddAuthor{I.~Atanassov}{}{}{a}%
	\AddAuthor{V.~Avati}{7}{}{}%
	\AddAuthor{J.~Baechler}{9}{}{}%
	\AddAuthor{C.~Baldenegro~Barrera}{11}{}{}%
	\AddAuthor{V.~Berardi}{4a}{4b}{}%
	\AddAuthor{M.~Berretti}{2a}{}{}%
	\AddAuthor{E.~Bossini}{6b}{}{}%
	\AddAuthor{U.~Bottigli}{6b}{}{}%
	\AddAuthor{M.~Bozzo}{5a}{5b}{}%
	\AddAuthor{H.~Burkhardt}{9}{}{}%
	\AddAuthor{F.~S.~Cafagna}{4a}{}{}%
	\AddAuthor{M.~G.~Catanesi}{4a}{}{}%
	\AddAuthor{M.~Csan\'{a}d}{3a}{}{b}%
	\AddAuthor{T.~Cs\"{o}rg\H{o}}{3a}{3b}{}%
	\AddAuthor{M.~Deile}{9}{}{}%
	\AddAuthor{F.~De~Leonardis}{4c}{4a}{}%
	\AddAuthor{M.~Doubek}{1c}{}{}%
	\AddAuthor{D.~Druzhkin}{9}{}{}%
	\AddAuthor{K.~Eggert}{10}{}{}%
	\AddAuthor{V.~Eremin}{}{}{c}%
	\AddAuthor{F.~Ferro}{5a}{}{}%
	\AddAuthor{A.~Fiergolski}{9}{}{}%
	\AddAuthor{F.~Garcia}{2a}{}{}%
	\AddAuthor{V.~Georgiev}{1a}{}{}%
	\AddAuthor{S.~Giani}{9}{}{}%
	\AddAuthor{L.~Grzanka}{7}{}{}%
	\AddAuthor{J.~Hammerbauer}{1a}{}{}%
	\AddAuthor{V.~Ivanchenko}{8}{}{}%
	\AddAuthor{J.~Ka\v{s}par}{6a}{1b}{}%
	\AddAuthor{J.~Kopal}{9}{}{}%
	\AddAuthor{V.~Kundr\'{a}t}{1b}{}{}%
	\AddAuthor{S.~Lami}{6a}{}{}%
	\AddAuthor{G.~Latino}{6b}{}{}%
	\AddAuthor{R.~Lauhakangas}{2a}{}{}%
	\AddAuthor{C.~Lindsey}{11}{}{}%
	\AddAuthor{M.~V.~Lokaj\'{\i}\v{c}ek}{1b}{}{}%
	\AddAuthor{L.~Losurdo}{6b}{}{}%
	\AddAuthor{M.~Lo~Vetere}{5b}{5a}{+}%
	\AddAuthor{F.~Lucas~Rodr\'{i}guez}{9}{}{}%
	\AddAuthor{M.~Macr\'{\i}}{5a}{}{}%
	\AddAuthor{M.~Malawski}{7}{}{}%
	\AddAuthor{N.~Minafra}{11}{}{}%
	\AddAuthor{S.~Minutoli}{5a}{}{}%
	\AddAuthor{T.~Naaranoja}{2a}{2b}{}%
	\AddAuthor{F.~Nemes}{9}{3a}{}%
	\AddAuthor{H.~Niewiadomski}{10}{}{}%
	\AddAuthor{T.~Nov\'{a}k}{3b}{}{}%
	\AddAuthor{E.~Oliveri}{9}{}{}%
	\AddAuthor{F.~Oljemark}{2a}{2b}{}%
	\AddAuthor{M.~Oriunno}{}{}{d}%
	\AddAuthor{K.~\"{O}sterberg}{2a}{2b}{}%
	\AddAuthor{P.~Palazzi}{9}{}{}%
	\AddAuthor{V.~Passaro}{4c}{4a}{}%
	\AddAuthor{Z.~Peroutka}{1a}{}{}%
	\AddAuthor{J.~Proch\'{a}zka}{1c}{}{}%
	\AddAuthor{M.~Quinto}{4a}{4b}{}%
	\AddAuthor{E.~Radermacher}{9}{}{}%
	\AddAuthor{E.~Radicioni}{4a}{}{}%
	\AddAuthor{F.~Ravotti}{9}{}{}%
	\AddAuthor{E.~Robutti}{5a}{}{}%
	\AddAuthor{C.~Royon}{11}{}{}%
	\AddAuthor{G.~Ruggiero}{9}{}{}%
	\AddAuthor{H.~Saarikko}{2a}{2b}{}%
	\AddAuthor{A.~Scribano}{6a}{}{}%
	\AddAuthor{J.~Smajek}{9}{}{}%
	\AddAuthor{W.~Snoeys}{9}{}{}%
	\AddAuthor{J.~Sziklai}{3a}{}{}%
	\AddAuthor{C.~Taylor}{10}{}{}%
	\AddAuthor{E.~Tcherniaev}{8}{}{}%
	\AddAuthor{N.~Turini}{6b}{}{}%
	\AddAuthor{V.~Vacek}{1c}{}{}%
	\AddAuthor{J.~Welti}{2a}{2b}{}%
	\AddAuthor{J.~Williams}{11}{}{}%
}


\def\DeclareInstitutes{%
	\AddInstitute{1a}{University of West Bohemia, Pilsen, Czech Republic.}
	\AddInstitute{1b}{Institute of Physics of the Academy of Sciences of the Czech Republic, Prague, Czech Republic.}
	\AddInstitute{1c}{Czech Technical University, Prague, Czech Republic.}
	\AddInstitute{2a}{Helsinki Institute of Physics, University of Helsinki, Helsinki, Finland.}
	\AddInstitute{2b}{Department of Physics, University of Helsinki, Helsinki, Finland.}
	\AddInstitute{3a}{Wigner Research Centre for Physics, RMKI, Budapest, Hungary.}
	\AddInstitute{3b}{EKU KRC, Gy\"ongy\"os, Hungary.}
	\AddInstitute{4a}{INFN Sezione di Bari, Bari, Italy.}
	\AddInstitute{4b}{Dipartimento Interateneo di Fisica di Bari, Bari, Italy.}
	\AddInstitute{4c}{Dipartimento di Ingegneria Elettrica e dell'Informazione - Politecnico di Bari, Bari, Italy.}
	\AddInstitute{5a}{INFN Sezione di Genova, Genova, Italy.}
	\AddInstitute{5b}{Universit\`{a} degli Studi di Genova, Italy.}
	\AddInstitute{6a}{INFN Sezione di Pisa, Pisa, Italy.}
	\AddInstitute{6b}{Universit\`{a} degli Studi di Siena and Gruppo Collegato INFN di Siena, Siena, Italy.}
	\AddInstitute{7}{AGH University of Science and Technology, Krakow, Poland.}
	\AddInstitute{8}{Tomsk State University, Tomsk, Russia.}
	\AddInstitute{9}{CERN, Geneva, Switzerland.}
	\AddInstitute{10}{Case Western Reserve University, Dept.~of Physics, Cleveland, OH, USA.}
	\AddInstitute{11}{The University of Kansas, Lawrence, USA.}
}

	
\def\DeclareExternalInstitutes{%
	\AddExternalInstitute{a}{INRNE-BAS, Institute for Nuclear Research and Nuclear Energy, Bulgarian Academy of Sciences, Sofia, Bulgaria.}
	\AddExternalInstitute{b}{Department of Atomic Physics, ELTE University, Budapest, Hungary.}
	\AddExternalInstitute{c}{Ioffe Physical - Technical Institute of Russian Academy of Sciences, St.~Petersburg, Russian Federation.}
	\AddExternalInstitute{d}{SLAC National Accelerator Laboratory, Stanford CA, USA.}
	\AddExternalInstitute{+}{Deceased.}
}



\newif\ifFirstAuthor
\FirstAuthortrue

\def\AddAuthor#1#2#3#4{%
	\def\PriAf{#2}%
	\def\SecAf{#3}%
	\def\ExtAf{#4}%
	\def\empty{}%
	\ifFirstAuthor
		\FirstAuthorfalse
	\else
		,
	\fi
	\ifx\PriAf\empty
		#1\Aref{#4}%
	\else
		\ifx\SecAf\empty
			\ifx\ExtAf\empty
				#1\Iref{#2}%
			\else
				#1\IAref{#2}{#4}%
			\fi
		\else
			\ifx\ExtAf\empty
				#1\IIref{#2}{#3}%
			\else
				#1\IIAref{#2}{#3}{#4}%
				\relax
			\fi
		\fi
	\fi
}


\def\AddCorrespondingAuthor#1#2#3#4#5#6{%
	\AddAuthor{#1}{#2}{#3}{*}%
	\Anotfoot{*}{#5 E-mail address: #6.}
}


\def\AddInstitute#1#2{%
	\expandafter\write1{\string\newlabel{#1}{{#1}{}}}%
	\hbox to\hsize{\strut\hss$^{#1}$#2\hss}%
}


\def\AddExternalInstitute#1#2{%
	\Anotfoot{#1}{#2}%
}



\begin{Authlist}
	\DeclareAuthors
\end{Authlist}

\DeclareInstitutes
\hbox to\hsize{\strut\hss} 
\DeclareExternalInstitutes

\date{\today}
\begin{abstract}
The proton-proton elastic differential cross section ${\rm d}\sigma/{\rm d}t$ has been measured by the TOTEM experiment at $\sqrt{s}=2.76$~TeV energy with $\beta^{*}=11$~m beam optics. The
Roman Pots were inserted to 13 times the transverse beam size from the beam, which allowed to measure the differential cross-section of elastic scattering in a range of the squared four-momentum transfer ($|t|$) from $0.36$~GeV$^{2}$ to $0.74$~GeV$^{2}$. The differential cross-section can be described with an exponential
in the $|t|$-range between $0.36$~GeV$^{2}$ and $0.54$~GeV$^{2}$, followed by a diffractive minimum (dip) at $|t_{\rm dip}|=(0.61\pm0.03)$~GeV$^{2}$ and a subsequent maximum (bump). The ratio of the ${\rm d}\sigma/{\rm d}t$ at the bump
and at the dip is $1.7\pm0.2$. When compared to the $\rm p\bar{p}$ measurement of the D0 experiment at $\sqrt s = 1.96\un{TeV}$, a significant difference can be observed. Under the condition
that the effects due to the energy difference between TOTEM and D0 can be neglected, the result provides evidence for a colourless 3-gluon bound state exchange in the $t$-channel of the proton-proton elastic scattering.

\end{abstract}

\end{titlepage}


\section{Introduction}

This article presents the first measurement of the proton-proton (pp) elastic differential cross section ${\rm d}\sigma/{\rm d}t$ at a centre-of-mass energy $\sqrt{s}=2.76$~TeV. The four momentum transfer squared ($|t|$) range
of the differential cross-section ${\rm d}\sigma/{\rm d}t$ includes the diffractive minimum. The TOTEM collaboration has previously measured proton-proton elastic scattering at energies 7~TeV, 8~TeV and 13~TeV~\cite{Antchev:2013paa,Antchev:2016vpy,Antchev:2011vs,Antchev:2013iaa,Nemes:2017gut,Antchev:2017dia,Paper_2p76}.
The importance of the present article is that it constitutes the pp ${\rm d}\sigma/{\rm d}t$ measurement closest to a corresponding $\rm p\bar{p}$ measurement at the TeV scale, since the D0 measurement is at a comparable energy $\sqrt s = 1.96$~TeV. The predominant Pomeron contribution to elastic pp scattering
is crossing even. Any difference between the pp and p$\bar{\rm p}$ differential cross-section at the TeV scale may be an evidence for a crossing-odd exchange, the Odderon, introduced in~\cite{Lukaszuk:1973nt,Gauron:1992zc} and
predicted in QCD as a 3-gluon bound state exchange~\cite{Bartels:1999yt}. At
the TeV energy scale, any possible other contribution by Reggeons is expected to be below the percent level~\cite{Jenkovszky:2017efs}.

Section~\ref{experimental_apparatus} outlines the experimental apparatus used for this measurement. Section~\ref{data_taking} summarises the data-taking conditions
including details of the kinematics reconstruction, alignment and beam optics. The differential cross-section is described in Section~\ref{differential_cross_section_section} followed
by a discussion of the physics results in Section~\ref{discussion_of_physics_results}.

\section{Experimental apparatus}
\label{experimental_apparatus}

The TOTEM experimental setup consists of two inelastic telescopes T1 and T2 to detect charged particles coming from inelastic $\rm pp$ collisions and
the Roman Pot detectors (RP) to detect elastically scattered protons at very small angles.
The inelastic telescopes are placed symmetrically on both sides of Interaction Point 5 (IP5): the T1
telescope is based on cathode strip chambers (CSCs)
placed at $\pm$9~m and covers the pseudorapidity range 3.1~$\le |\eta| \le$~4.7; the T2 telescope is based on gas electron
multiplier (GEM) chambers placed at $\pm$13.5~m and covers the pseudorapidity range 5.3~$\le |\eta| \le$~6.5. The pseudorapidity coverage of the two telescopes at $\sqrt{s}=2.76$~TeV allows the detection of about 92~\% of the inelastic events. As the fraction of events with all final state particles beyond the instrumented region
has to be estimated using phenomenological models, the	
excellent acceptance in TOTEM minimizes the dependence on such models and thus provides small uncertainty on the
inelastic rate measurement.

The Roman Pot (RP) units used for the present measurement are located on both sides of the IP at distances of $\pm214.6$~m (near) and $\pm220.0$~m (far) from IP5. A unit consists of 3 RPs, two approaching the outgoing beam vertically and one horizontally.
The horizontal RP detectors were not inserted during this particular data taking and the vertical alignment uses the RP position sensors and is further refined with precise constraints based on symmetries of elastic scattering~\cite{Nemes:2017gut}.
The $5.4$~m long lever arm between the near and the far RP units has the important advantage that the local track angles in the $x$ and $y$-projections perpendicular to the beam direction can be reconstructed with a precision of 2~$\mu$rad. A complete description of the TOTEM detector
system is given in~\cite{Anelli:2008zza,TOTEM:2013iga}.

			Each RP is equipped with a stack of 10 silicon strip detectors designed with the specific objective of reducing the insensitive area at the edge facing the beam to only a few tens of micrometres. The 512 strips with 66
			$\mu$m pitch of each detector are oriented at an angle of +45$^{\circ}$ (five planes) and -45$^{\circ}$ (five planes) with respect to the
			detector edge facing the beam~\cite{Ruggiero:2009zz}.

	\vspace{1mm}
	\begin{figure}[H]
		\centering
		\includegraphics[trim = 0mm 510mm 610mm 0mm, clip, width=1.0\columnwidth]{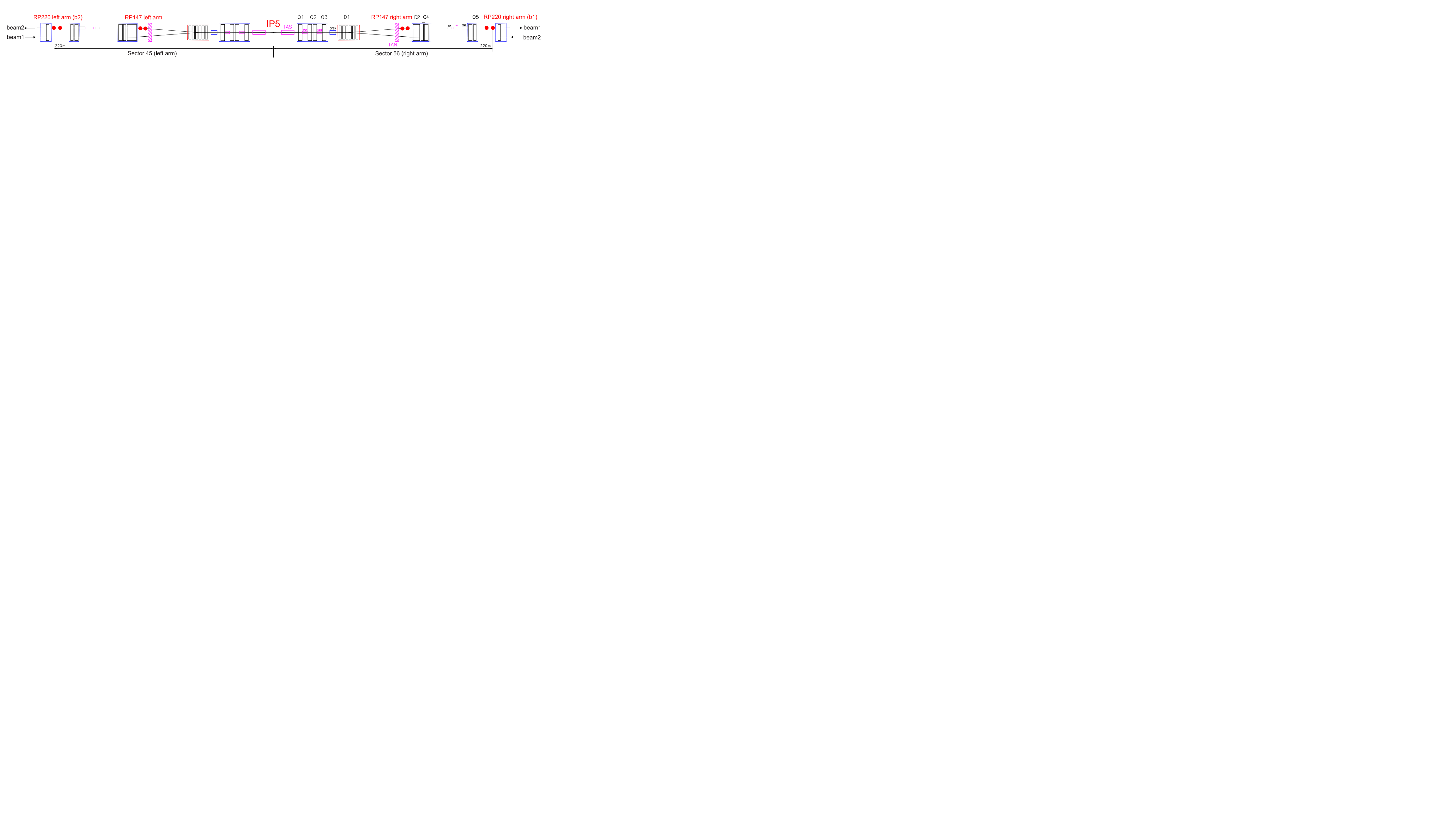}
		\caption{(color) Schematic layout of the LHC from IP5 up to the “near” and “far” Roman Pot units, where the near and far pots are indicated by full (red) dots on beams 1 and 2.}
		\label{layout_LHC}
 	\end{figure}
	\vspace{2mm}

\section{Data taking and analysis}
\label{data_taking}

The analysis is performed on a data sample (DS1) recorded in 2013 during an LHC fill with $\beta^{*}=11$~m injection optics~\cite{Bruning:2004ej,Antchev:2014voa,Nemes:2131667}. The RP detectors
were inserted to 13 times the transverse beam size. Although in this work we focus on the analysis of DS1, the present analysis uses the total cross-section measurement at $\sqrt{s}=2.76$~TeV (data set DS2) recorded with RP detectors placed at 4.3 times the transverse beam size, in order to obtain its final normalization~\cite{Antchev:2017dia}.
The differential cross-section of DS2 is included in this article for the sake of completeness, see Table~\ref{DS2_data}, although its detailed description is provided elsewhere~\cite{Antchev:2017dia,Paper_2p76}.

The vertical RP detectors were at 13 times the transverse beam size ($\sigma_{\rm beam}$) from the outgoing beams. The collected events have been triggered by the T2
telescope in either arm (inelastic trigger), by the RP detectors in a double-arm coincidence (elastic trigger), and by
random bunch crossings (zero-bias sample used for calibration).

	\subsection{Elastic analysis}

	\subsubsection{Reconstruction of kinematics}
		The horizontal and vertical scattering angles of the proton at IP5 $(\theta_{x}^{*},\theta_{y}^{*})$  are reconstructed in a given arm by inverting the proton transport
		equations~\cite{Antchev:2014voa}
			\begin{align}
			    \theta_{x}^{*} = \frac{1}{\frac{{\rm d}L_{x}}{{\rm d}s}}\left(\theta_{x}-\frac{{\rm d} v_{x}}{{\rm d} s}x^{*}\right)\,,\,\,\,
    			    \theta^{*}_{y} = \frac{y}{L_{y}}\,,
			    \label{reconstruction_formula_theta_x_rearranged}
			\end{align}
			where $s$ denotes the distance from the interaction point, $y$ is the vertical coordinate of the proton's trajectory, $\theta_{x}$ is its horizontal
			angle measured by the RP detectors, and $x^{*}$ is the horizontal vertex coordinate reconstructed as
			\begin{align}
			    x^{*}&=\frac{L_{x,{\rm far}}\cdot x_{\rm near} - L_{x,{\rm near}}\cdot x_{\rm far}}{d}\,,
			    \label{reconstruction_formula_x_star}
			\end{align}
			where $d=( v_{x,{\rm near}}\cdot L_{x,{\rm far}} -  v_{x,{\rm far}}\cdot L_{x,{\rm near}})$. The coefficients $L_{x}$, $L_{y}$ and $v_{x}$ are optical functions of the LHC beam determined by the
			accelerator magnets. For their definition we refer to~\cite{Antchev:2014voa}.

			The scattering angles obtained for the two arms are averaged and the four-momentum transfer squared is calculated
			\begin{align}
			    t=-p^{2}\theta^{*2}\,,
			    \label{reconstructed_t}
			\end{align}
		where $p=1.38$~TeV is the LHC beam momentum and the scattering angle $\theta^{*}=\sqrt{{\theta_{x}^{*}}^{2} + {\theta_{y}^{*}}^{2}}$. Finally, the azimuthal angle is 
			\begin{align}
				\phi^{*}=\arctan\left(\frac{\theta_{y}^{*}}{\theta_{x}^{*}}\right)\,.
			    \label{phistar}
			\end{align}

	\subsubsection{RP alignment and beam optics}
	\label{RP_alignment}

	The alignment is based on the position measurement of the RP movement system, followed by an alignment procedure based on the symmetries of elastic scattering. The residual misalignment with respect to the LHC beam is about
	10~$\mu$m in the horizontal coordinate and about 100~$\mu$m in the vertical~\cite{Antchev:2016vpy,Antchev:2015zza}. When propagated to the reconstructed scattering angle $\theta^{*}$, this leads to an uncertainty of
	the order 5~$\mu$rad.

	The nominal optics has been updated from LHC magnet and current databases and has been calibrated using the observed elastic candidates of DS2, with larger statistics, and validated for
	DS1 relying on the stability of the LHC optics~\cite{Bruning:2004ej,Nemes:2017gut}.  The $\beta^{*}=11$~m optics of the LHC is designed with a vertical effective length $L_{y}\approx~19.4$~m at the location of the RP detectors;
	the exact value depends on the location of the detector along the beam. The reconstruction
		of the horizontal scattering angle uses the derivative of the horizontal effective length ${{\rm d}L_{x}}/{{\rm d}s}\approx-0.4$ at the position of the RPs. The remaining optical functions used
		in the reconstruction are the horizontal magnifications in the near and far RP, whose value is $v_{x, {\rm near}} \approx v_{x, {\rm far}}\approx-3.2$ and their derivative ${{\rm d}v_{x}}/{{\rm d}s}\approx 4.9\cdot10^{-2}~$m$^{-1}$. The different reconstruction formula in the vertical
		and horizontal plane in Eq.~(\ref{reconstruction_formula_theta_x_rearranged}) is motivated by their different sensitivity to the LHC magnet and beam perturbations.

	\begin{figure}[H]
		\centering
		\includegraphics[width=0.48\columnwidth,height=52mm]{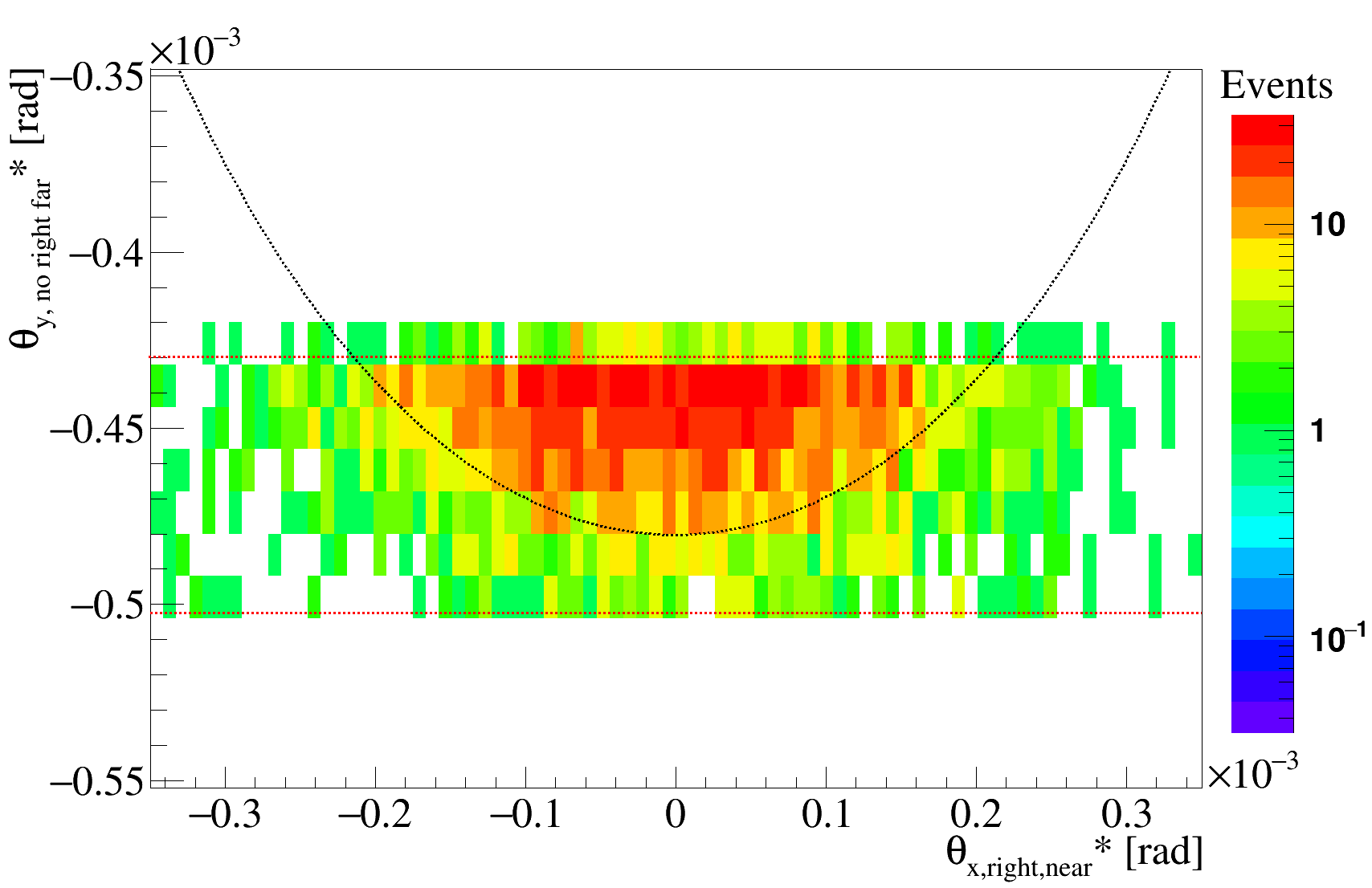}
		\includegraphics[width=0.48\columnwidth,height=50mm]{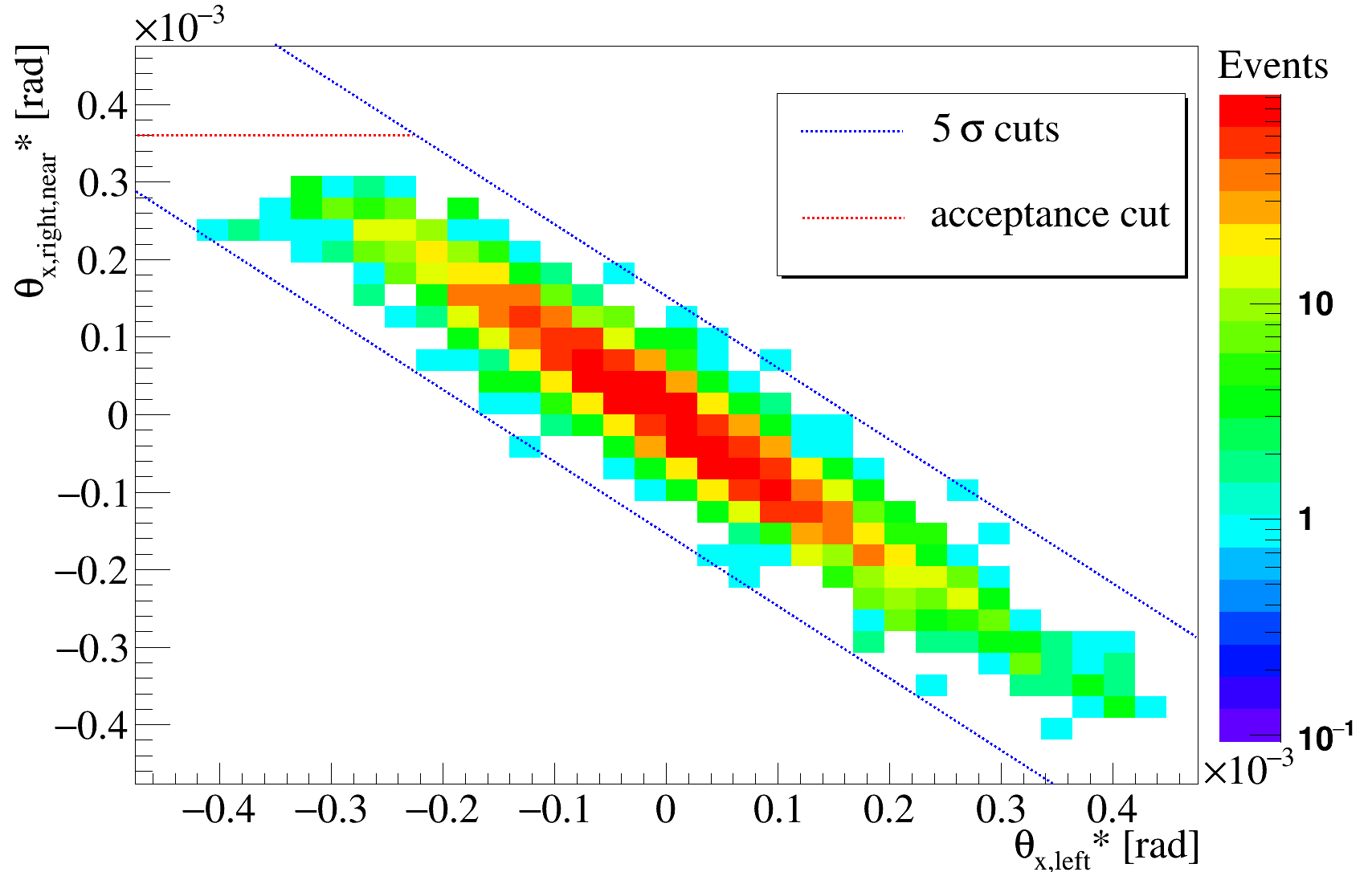}
		\caption{(color) Geometrical acceptance cut in Diagonal 1 on the $(\theta_{x}^{*},\theta_{y}^{*})$ plane (left panel) and the collinearity cut of the two protons using the
		horizontal scattering angle $\theta_{x}^{*}$ (right panel). The red and blue lines show the acceptance and 5$\sigma$ physics cuts, respectively. In order to optimize the
		acceptance the right far RP was not used, denoted in $\theta^{*}_{y,\rm no~right~far}$, see also Section~\ref{event_selection}.}
		\label{cuts}
 	\end{figure}

	The uncertainties of the optical functions are estimated with a Monte Carlo program applying the optics calibration procedure
	on a sophisticated simulation of the LHC beam and its perturbations. The obtained uncertainty is  2~$\permil$ for ${{\rm d}L_{x}}/{{\rm d}s}$ and $3~\permil$ for $L_{y}$.
	The uncertainty of the horizontal magnification $v_{x}$ and its derivative is 2 and 3~$\permil$, respectively~\cite{Antchev:2014voa,Nemes:2131667}.

	\begin{figure}[H]
		\centering
		\includegraphics[width=0.68\columnwidth]{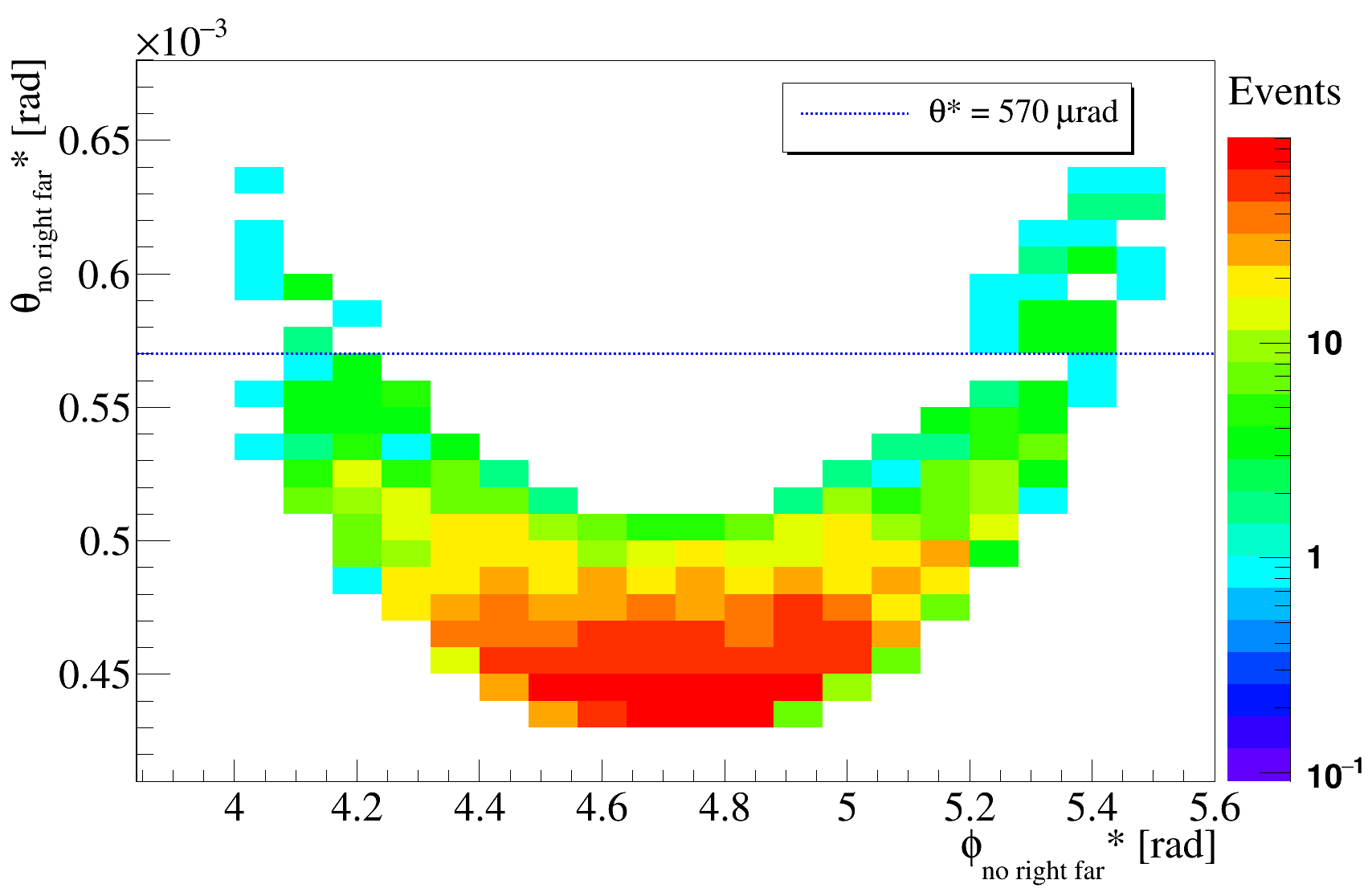}
		\caption{(color) The uncorrected distribution of the azimuthal angles $\phi^{*}$ per event as a function of $\theta^{*}$ in Diagonal 1. The consecutive empty bins along the $\theta^{*}=570$~$\mu$rad
		line are due to the diffractive minimum at $t\approx-0.61$~GeV$^{2}$, see Figure~\ref{differential_cross_section}. In order to optimize the
		acceptance the right far RP was not used, denoted in $\phi^{*}_{\rm no~right~far}$, see also Section~\ref{event_selection}.}
		\label{cuts2}
 	\end{figure}

	\subsubsection{Event selection}
	\label{event_selection}

	The analysis follows the similar procedure used for the measurement of the elastic cross section at several other LHC energies:~7~TeV, 8~TeV and 13~TeV~\cite{Antchev:2013paa,Antchev:2016vpy,Antchev:2011vs,Antchev:2013iaa,Nemes:2017gut,Antchev:2017dia,Paper_2p76}. 
	The measurement of the elastic rate is based on the selection of events with the following topology in the RP detector system: a reconstructed track in the near and far
	vertical detectors on one side and a reconstructed track in the near (or far) on the other side of the IP such that the elastic signature is satisfied in one of the two diagonals: left bottom and right top (Diagonal 1) or left top and right bottom (Diagonal 2).
	
	There are four vertical RP detectors along a diagonal, each with slightly different acceptance limitations depending on their vertical distance from the LHC beam. The mentioned topology selection uses only three RPs in order to optimize the statistics of the analysis.
	In the arm with only one RP the horizontal scattering angle is reconstructed using
			\begin{align}
			    \theta_{x}^{*} = \frac{1}{L_{x}}\left(x- v_{x}\cdot x^{*}\right)\,,
			    \label{reconstruction_formula_theta_x_rearranged_from_x}
			\end{align}
	where the horizontal vertex coordinate $x^{*}$ is calculated from Eq.~(\ref{reconstruction_formula_x_star}) using the track of the RPs in the other arm of the diagonal.
	
\begin{figure}[H]
		\centering
		\includegraphics[width=0.85\columnwidth]{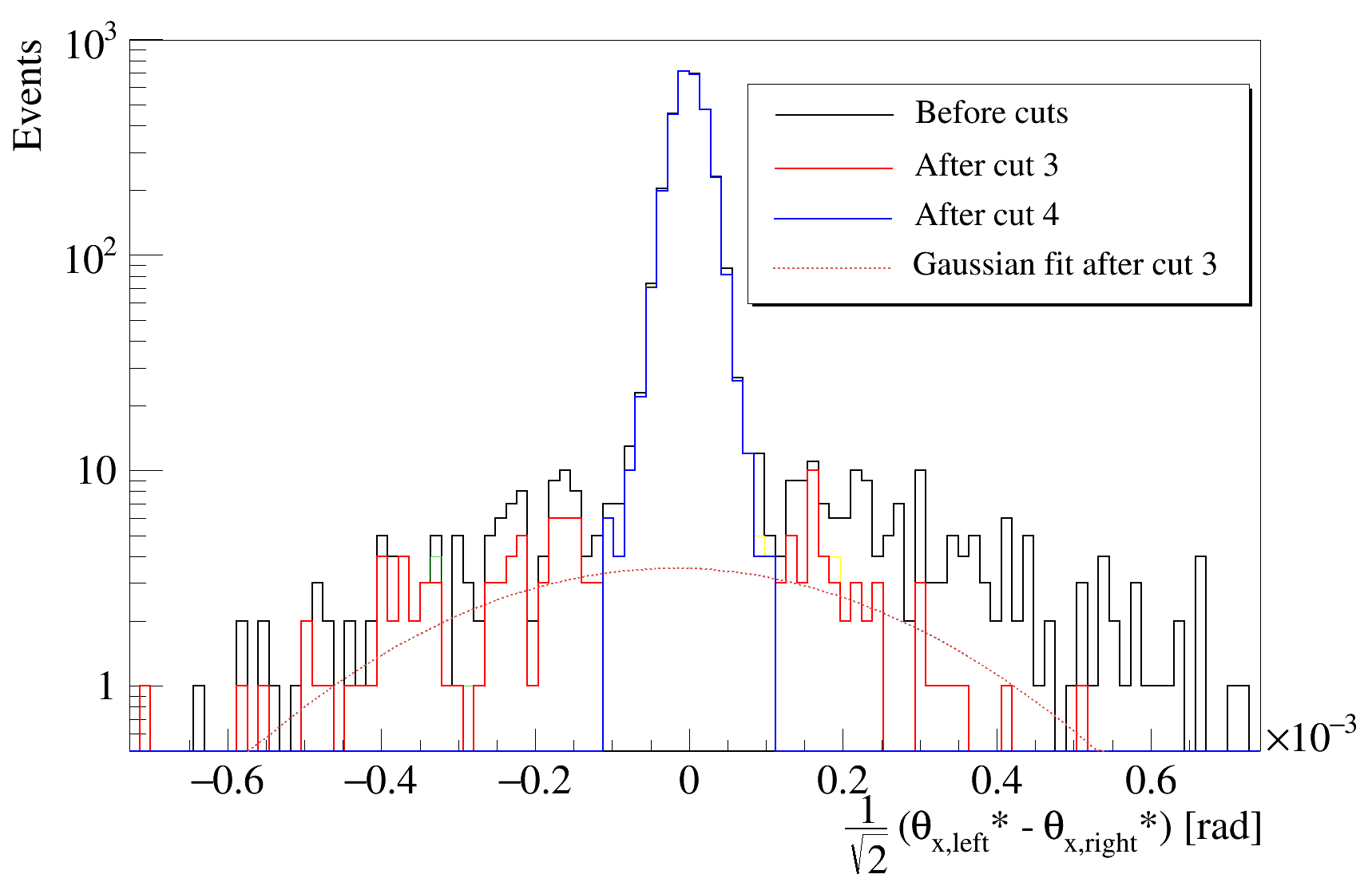}
		\caption{(color) The horizontal beam divergence estimated from the data in Diagonal 1. The distribution is shown before
		any analysis cut (black solid line) and before and after the last cut, see Table~\ref{cuts_table}. The residual background is estimated with a Gaussian fit of the tail before
		the last analysis cut. }
		\label{signal_to_noise}
	\end{figure}

	Besides the topology cut, the elastic event selection requires the collinearity of the outgoing protons in the two arms, see Fig.~\ref{cuts}. The diffractive events are suppressed
	with so-called spectrometer cuts, which require the correlation between the vertical position in the near RP detector $y_{\rm near}$ and the inclination $\Delta y=y_{\rm far}-y_{\rm near}$ (and similarly in the horizontal plane), see Table~\ref{cuts_table}. Fig.~\ref{signal_to_noise} shows the efficiency of the elastic event selection.

	\begin{table*}\small\color{black}
        \begin{center}
            \caption{(color) The physics analysis cuts and their width $\sigma$ in Diagonal 1 (the other diagonal is in agreement within the quoted uncertainty). The width $\sigma$ of the horizontal and vertical collinearity cuts define the resolution in the scattering angle, see Figure~\ref{cuts}.}
            \begin{tabular}{ | c  l  c |}
		\hline
                		& \,\,\,\,\,\,\,\,Cut name	& 		$\sigma$		\\\hline 
			1	& Vertical collinearity cut	&  	 \,\,\,$21.3\pm0.4~\mu$rad		\\ 
			2	& y-spectrometer cut, left arm	& 	 $51.1\pm0.4~\mu$m		\\ 
			3	& x-spectrometer cut, left arm	& 	 $69.3\pm1.2~\mu$m		\\ 
			4	& Horizontal collinearity cut	& 	 \,\,\,$22.3\pm0.5~\mu$rad		\\\hline 
            \end{tabular}
        \label{cuts_table}
        \end{center}
    \end{table*}

	Figure~\ref{cuts} shows the horizontal collinearity cut imposing momentum conservation in the horizontal plane. The cuts are applied at the 5$\sigma$ level, and they are optimized for
	purity (background contamination in the selected sample	less than {0.5~\%)} and for efficiency (uncertainty of true elastic event selection 0.5~\%). Figure~\ref{signal_to_noise} shows the progressive selection
	of elastic events. 

	\begin{figure}[H]
		\centering
		\includegraphics[width=0.95\columnwidth]{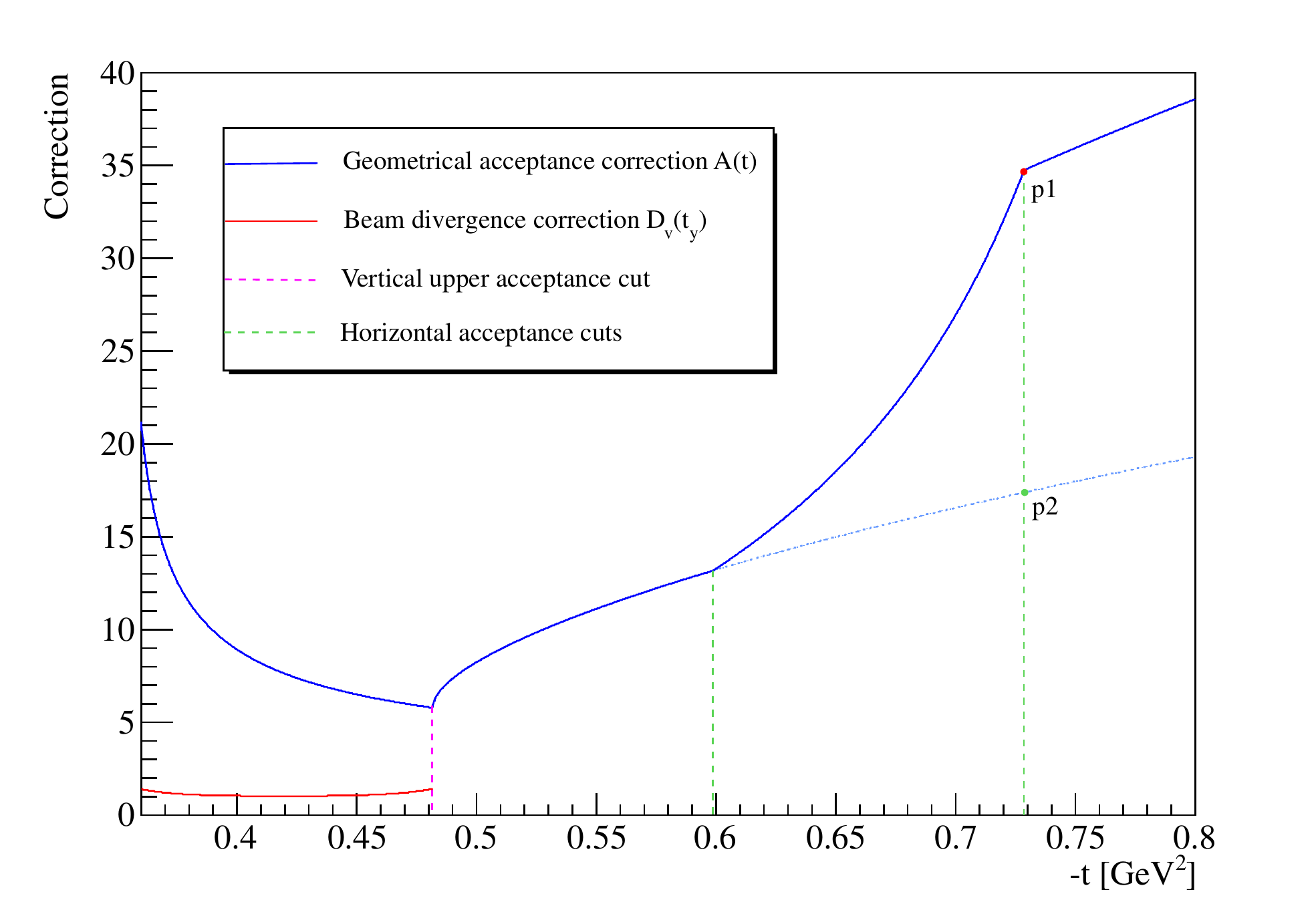}
		\caption{(color) The geometrical acceptance correction $A(t)$ and the vertical beam divergence correction $\mathcal{D}_{\rm v}(t_{y})$ for Diagonal 1. The vertical lines indicate the $|t|$-positions where the additional acceptance limitations appear due to the
		 vertical and horizontal LHC apertures. The dashed blue line indicates a hypothetical geometrical correction without horizontal acceptance cuts. The ordinate of point $p1$ is two times the ordinate of $p2$
		at the upper horizontal cut, since the acceptance is halved by the cut at the LHC aperture.}
		\label{geometrical_acceptance_correction}
	\end{figure}

	\subsubsection{Geometrical and beam divergence correction, unfolding}
		{\color{black} The vertical acceptance of elastically scattered protons is limited by the RP silicon detector edge and by the LHC magnet apertures. The geometrical acceptance correction is calculated in order
		to correct for the missing part of the acceptance
		\begin{equation}
			\mathcal{A}(\theta^{*})=\frac{2\pi }{\Delta \phi^{*}(\theta^{*})}\,,
			\label{geom_accep_formula}
		\end{equation}
 		where $\Delta \phi^{*}$ is the visible azimuthal angle range, defined by the acceptance cuts, see Figure~\ref{cuts}.

		The geometrical acceptance correction formula Eq.~(\ref{geom_accep_formula}) assumes the azimuthal symmetry of elastic scattering, which is experimentally verified on the data, see Fig.~\ref{cuts2}. The acceptance limitations constrain the vertical component $t_{y}$ of the analysis to
		{$|t|_{y,\rm min}=0.36$~GeV$^{2}$ and $|t|_{y,\rm max}=0.48$~GeV$^{2}$}. The RP distance from the LHC beam is larger than in the earlier TOTEM analyses and the geometrical acceptance correction
		factor $\mathcal{A}(t)$ exceeds 5.

		The scattering angles are large and reach LHC apertures horizontally. In
		both diagonals this angular cut has been measured per arm with a dedicated high statistics single arm analysis using 2 RPs. The tighter angular cut for Diagonal 1 is
		in the right arm $\theta_{x,\rm coll}^{*}=+360$~$\mu$rad, which is taken into account in the geometrical acceptance correction $\mathcal{A}(t)$, see Figure~\ref{geometrical_acceptance_correction}.
		The same procedure is applied for Diagonal 2. Figure~\ref{geometrical_acceptance_correction} also provides a reference curve for $\mathcal{A}(t)$ shown as a blue dashed line without the $\theta_{x,\rm coll}^{*}$
		cut. 

 		Close to the acceptance edges there is an additional acceptance loss due to the angular smearing, divergence, of the beam. This additional acceptance loss is modeled with a Gaussian distribution at
		the corners of the acceptance, with experimentally determined parameters. The model permits to calculate the corresponding vertical beam divergence correction $\mathcal{D}_{\rm v}(t_{y})$, see Figure~\ref{geometrical_acceptance_correction}. In the horizontal plane the beam divergence correction $\mathcal{D}_{\rm h}(t_{x})$ is below 0.5~\%.
	\begin{figure}[H]
		\centering
		\includegraphics[width=0.95\columnwidth]{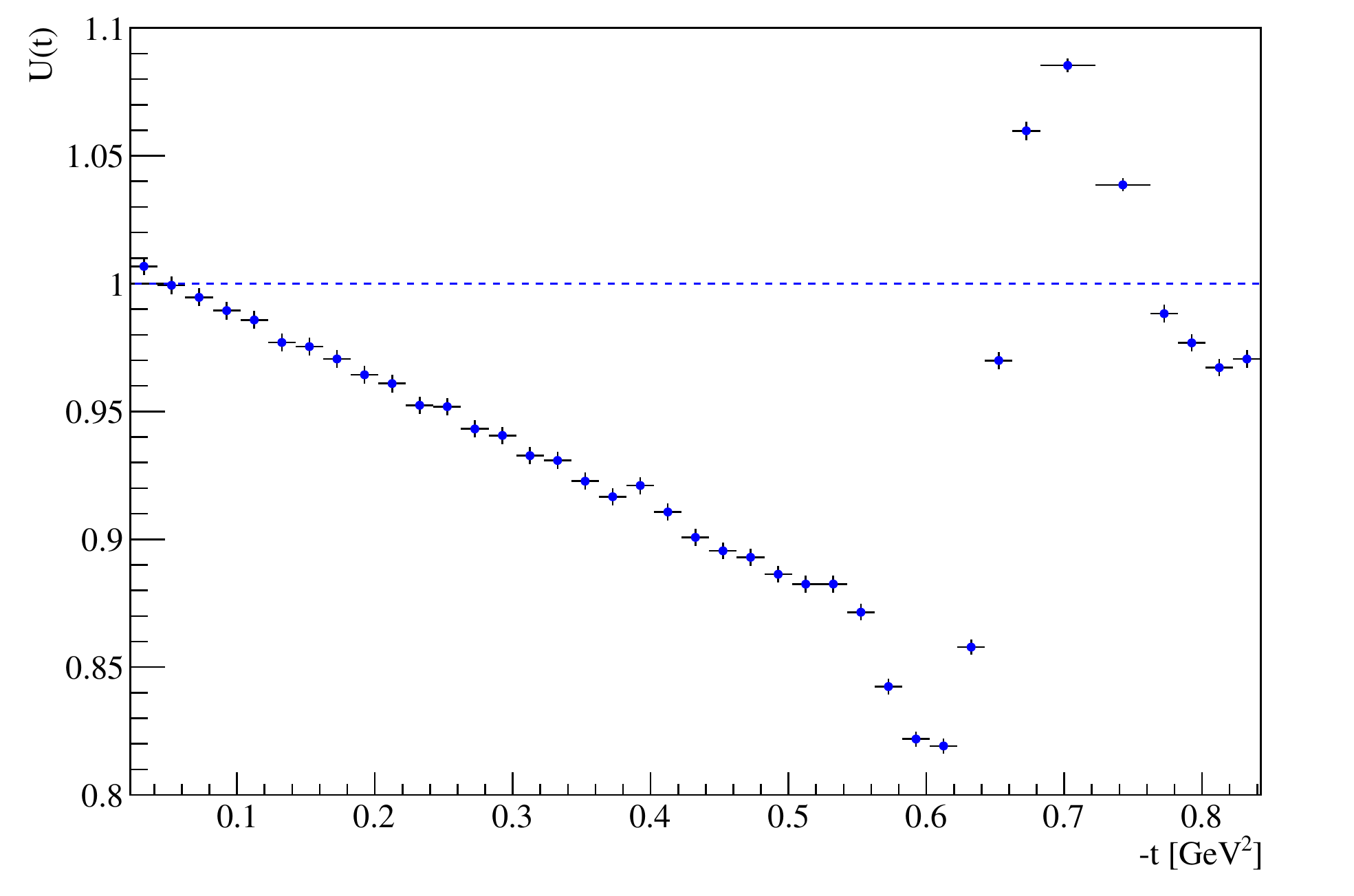}
		\caption{(color) The multiplicative unfolding correction histogram, obtained with the resolutions of the  vertical and horizontal scattering angle described in Table~\ref{cuts_table}. The magnitude of the
		correction reaches its maximum -18~\% around the dip, where the nuclear slope $B(t)$ changes the most. After the minimum the correction changes sign following the derivative of 
		$B(t)$ and just after $|t|=0.7$~GeV$^{2}$ the correction reaches +9~\%. }
		\label{unfolding_correction}
	\end{figure}

		The unfolding of resolution effects is estimated with a Monte Carlo simulation. The resolution parameters are obtained from the data, see Section~\ref{RP_alignment}. The probability distribution $p(t)$ of the
		event generator is based on the fit of the differential rate ${\rm d}N_{\rm el}/{\rm d}t$. Each generated MC event is propagated to the RP detectors with the proper model of the LHC optics, which takes
		into account the beam divergence and other resolution effects. The kinematics of the event is reconstructed and a histogram is built from
		the $t$ values. The ratio of the histograms without and with resolution effect describes the first approximation of the bin-by-bin corrections due to bin migration. The probability distribution
		$p(t)$ of the simulation is multiplied with the correction histogram, to modulate the source, and the procedure is repeated until the histogram with migration effects
		coincide with the measured distribution, thus the correct source distribution has been found. The uncertainty of the unfolding procedure is estimated from the residual difference between the measured histogram
		${\rm d}N_{\rm el}/{\rm d}t$ and the simulated histogram with resolution effects.

		The angular spread of the beam is determined with an
		uncertainty 0.5~$\mu$rad by comparing the scattering angles reconstructed from the left and right arm, see Table~\ref{cuts_table}. Therefore 
    		the unfolding correction factor $\mathcal{U}(t)$ can be calculated with a precision better than 0.1~\%}, see Figure~\ref{unfolding_correction}.
 		The event-by-event correction factor due to acceptance corrections and resolution unfolding is
		\begin{equation}
			\mathcal{C}(t_{x},t_{y})=\mathcal{A}(t)\mathcal{D}_{\rm v}(t_{y})\mathcal{D}_{\rm h}(t_{x})\mathcal{U}(t)\,.
		\end{equation}

\section{The differential cross section}
\label{differential_cross_section_section}
	The inefficiency corrections due to pile-up from background and inefficiency due to an additional
	inefficiency of one RP out of the three used is taken into account with a relative scale factor, computed as a ratio of Diagonal 1 to Diagonal 2 in a representative $|t|$-range. 

	After these corrections the differential rate ${\rm d}N_{\rm el}/{\rm d}t$ of Diagonal 1 and Diagonal 2 agree within their statistical uncertainty over the whole $|t|$-range measured. The two diagonals are almost
	independent measurements, thus the final measured differential rate is calculated as the bin-by-bin weighted average of the two differential elastic rates ${\rm d}N_{\rm el}/{\rm d}t$, according to their statistical uncertainty.

	\begin{figure}[H]
		\includegraphics[trim = 0mm 0mm 0mm 0mm, clip,width=0.9\linewidth]{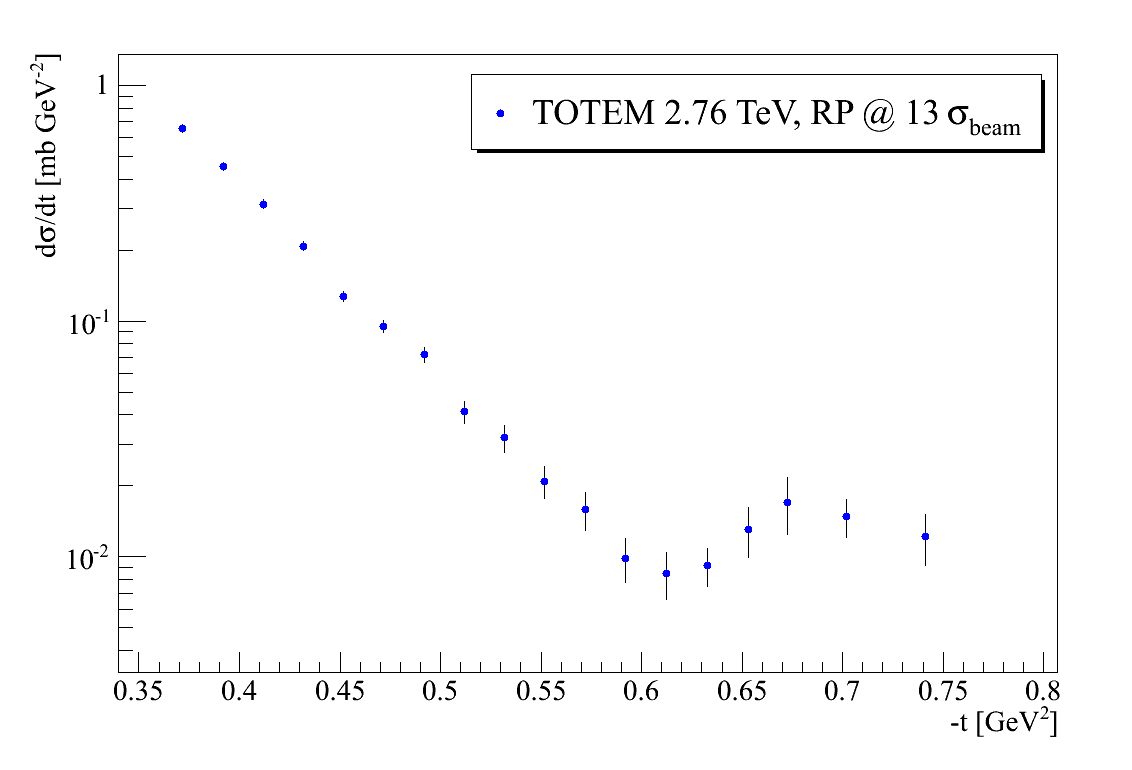}
		\caption{(color) The pp differential elastic cross section ${\rm d}\sigma/{\rm d}t$ of DS1 at $\sqrt{s}=2.76$~TeV.}
		\label{differential_cross_section}
	\end{figure}
	
\begin{figure*}[h]
		\includegraphics[trim = 0mm 0mm 0mm 0mm, clip,width=0.9\linewidth]{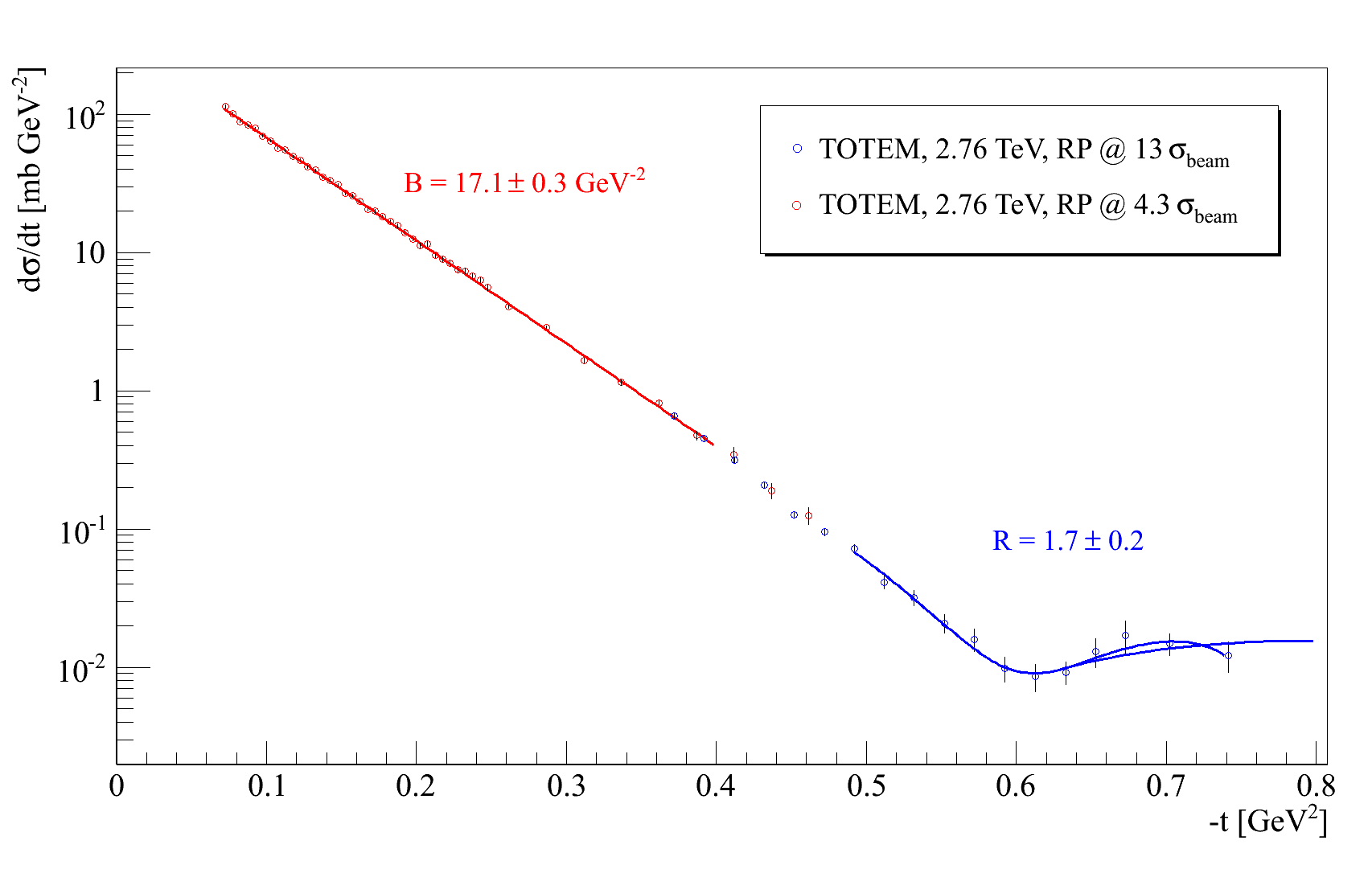}
		\caption{(color) The differential cross section ${\rm d}\sigma_{\rm el}/{\rm d}t$ at $\sqrt{s}=2.76$~TeV. The figure shows the dataset DS1 (blue hollow circles) and the dataset DS2
		of the total cross-section measurement (red hollow circles) used for normalization~\cite{Antchev:2017dia,Paper_2p76,Nemes:2017gut}. The nuclear slope $B=(17.1\pm0.3)$~GeV$^{-2}$ and the corresponding fit in the $|t|$-range between 0.09~GeV$^{2}$ and 0.4~GeV$^{2}$ is shown. The
		fit of the diffractive minimum and maximum with a third order polynomial, for two possible functional forms, is presented in the $|t|$-range between 0.47~GeV$^{2}$ and 0.74~GeV$^{2}$ and beyond.}
		\label{differential_cross_section_normalization}
	\end{figure*}

	The overall normalization is determined from the total cross-section analysis at $\sqrt{s}=2.76$~TeV, summarized in~\cite{Antchev:2017dia,Paper_2p76}. The final differential cross-section
	${\rm d}\sigma/{\rm d}t$ is obtained by normalizing DS1 to DS2 using the integral of their exponential in the overlapping $t$-range. The uncertainty on the normalization is about 6~\%. The
	differential cross-section of DS1 is shown in Figure~\ref{differential_cross_section}. Figure~\ref{differential_cross_section_normalization} shows the complete range of the differential cross-section
	covered by DS1 and DS2.
	The ${\rm d}\sigma/{\rm d}t$ data points are summarized in Table~\ref{DS1_data}, where the $|t|$-dependent systematic uncertainty is also provided. Table~\ref{DS2_data} contains
	the data points for DS2.

 	Fig.~\ref{differential_cross_section_normalization} shows the fit of the diffractive minimum and the possible positions of the subsequent maximum with a third order polynomial in the $|t|$-range between 0.47~GeV$^{2}$ and 0.74~GeV$^{2}$ and beyond. In fact
	the data determine and characterize the $t$-position of the dip $t_{\rm dip}$, the cross-section at $t_{\rm dip}$ and the cross-section at the bump (the local maximum subsequent to the dip) or a lower limit of such cross-section. However, the data do not constrain the $t$-position of the bump $t_{\rm bump}$ , which 
	could be anywhere in the range 0.7~--~0.8~GeV$^{2}$ without effecting the corresponding cross-section given the flat derivative (Fig.~\ref{differential_cross_section_normalization}). The dip position is found to be
	$|t_{\rm dip}|=(0.61\pm0.03)$~GeV$^{2}$. The overall uncertainty in $|t|$ (correlated bin-to-bin) is derived from the beam divergence (5~\%), alignment (less than 2~\%) and unfolding (less than 0.5~\%). 

	The nuclear slope $B=(17.1\pm0.3)$~GeV$^{-2}$ and the corresponding exponential fit in the $|t|$-range between 0.09~GeV$^{2}$ and 0.4~GeV$^{2}$ of the differential cross-section are perfectly consistent with~\cite{Paper_2p76,Antchev:2017dia}.
	
\begin{figure*}[h]
		\includegraphics[trim = 0mm 0mm 0mm 0mm, clip,width=0.9\linewidth]{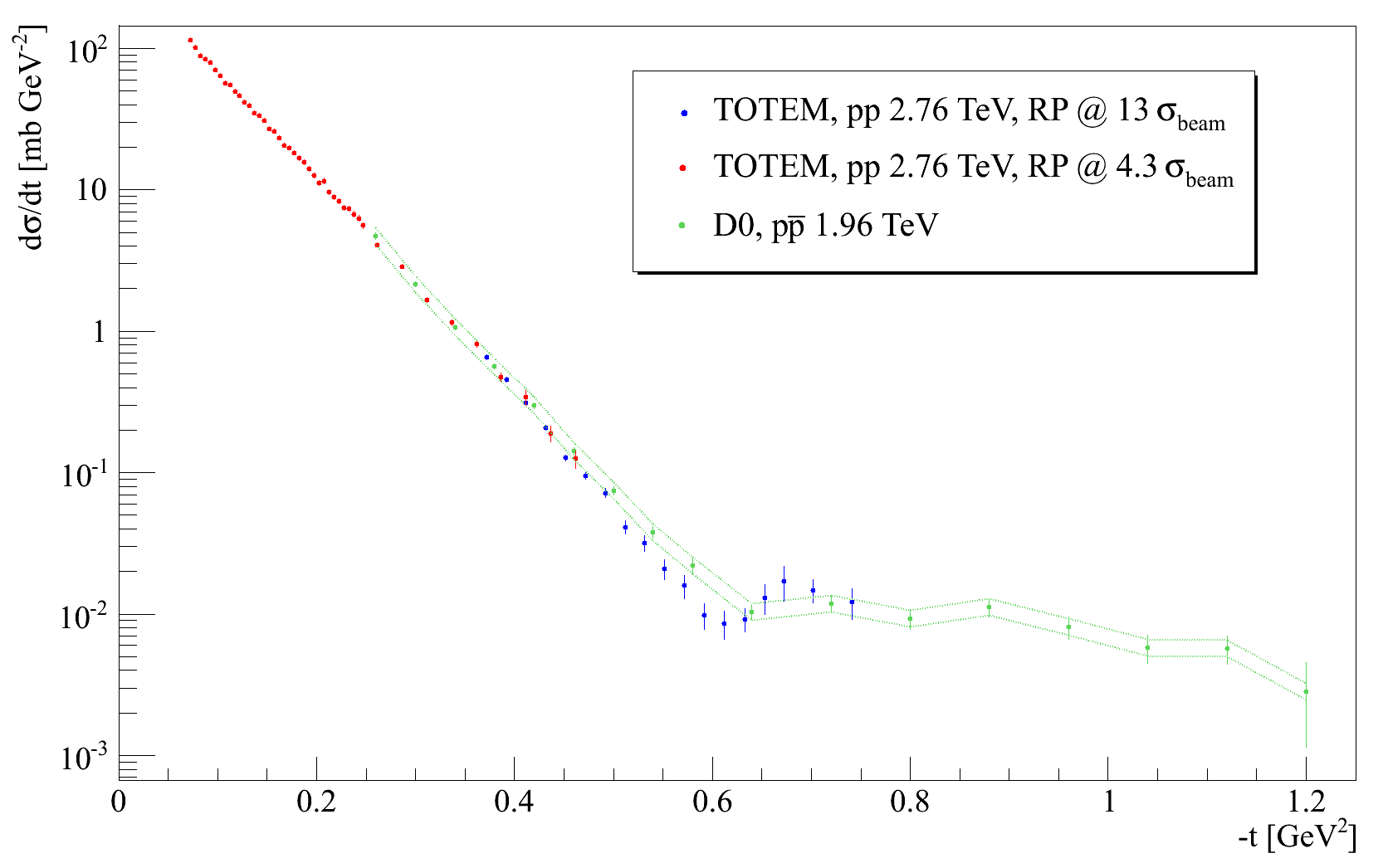}
		\caption{(color) The differential cross sections ${\rm d}\sigma/{\rm d}t$ at $\sqrt{s}=2.76$~TeV measured by the TOTEM experiment and the elastic $\rm p\bar{p}$ measurement of the D0 experiment at 1.96 TeV~\cite{Abazov:2012qb}. The
		green dashed line indicates the normalization uncertainty of the D0 measurement.}
		\label{differential_cross_section_with_D0}
	\end{figure*}

\begin{figure*}[h]
		\includegraphics[trim = 0mm 0mm 0mm 0mm, clip,width=0.9\linewidth]{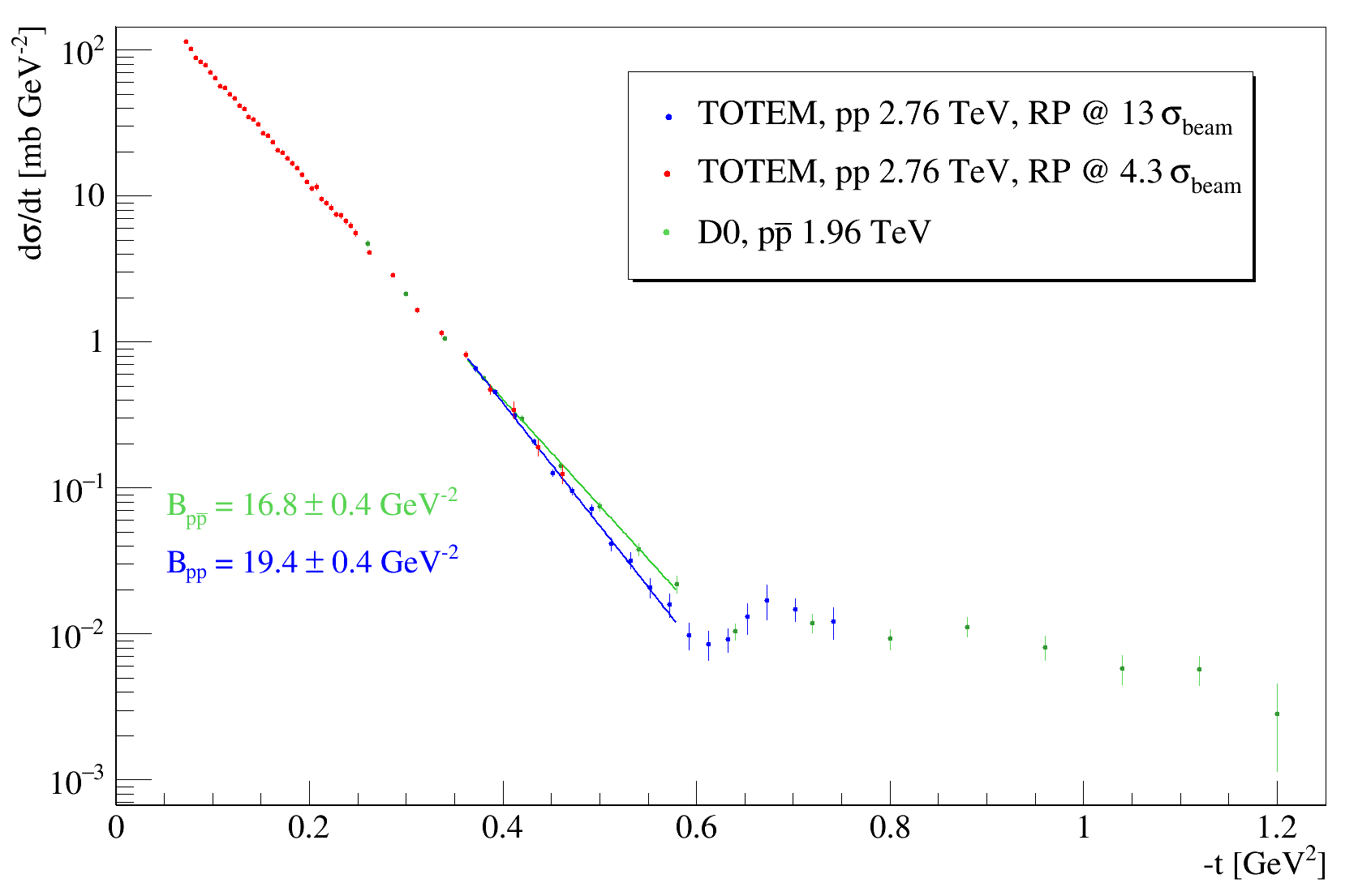}
		\caption{(color) The exponential fit of the differential cross sections ${\rm d}\sigma/{\rm d}t$ at $\sqrt{s}=2.76$~TeV measured by the TOTEM experiment and the elastic $\rm p\bar{p}$ measurement
		of the D0 experiment at 1.96 TeV in the $|t|$-range from 0.36~GeV$^{2}$ to 0.58~GeV$^{2}$. The pp differential cross section shows a steepening before the dip, and the slope parameters
		in this range quantify another key parameter to claim the significant deviation of pp and p$\rm \bar{p}$.}
		\label{differential_cross_section_with_D0_fits}
	\end{figure*}

	The differential cross-section ${\rm d}\sigma/{\rm d}t$ is compared to the p$\bar{\rm p}$ measurement of the D0 experiment in Figure~\ref{differential_cross_section_with_D0}.
 	The measured nuclear slopes before the dip $B_{\rm pp}=(19.4\pm0.4)$~GeV$^{-2}$ and $B_{\rm p\bar{p}}=(16.8\pm0.4)$~GeV$^{-2}$ are key parameters to quantify the difference between pp and p$\bar{\rm p}$, see Fig.~\ref{differential_cross_section_with_D0_fits}. According to
	the nuclear slope difference, the significance of the incompatibility between the pp vs. p$\rm\bar{p}$ is greater than 4$\sigma$. Recently, refs.~\cite{Martynov:2018sga,Csorgo:2018uyp}, pointed out
	that the $t$-dependent nuclear slope parameter $B(t)=\frac{\rm d}{{\rm d}t}\ln({\rm d}\sigma/{\rm d}t)$ indicates a clear Odderon effect as $B_{\rm pp}(t) \ne B_{\rm p\bar{p}}(t)$.

	A dedicated Monte Carlo simulation has been used to simulate all analysis steps in order to model the correct propagation of the central values and their uncertainties.
	The simulation resulted in uncertainty corrections mainly due to the asymmetry of Poisson distributions	in the bins which have lower statistics. The uncertainty on the
	${\rm d}\sigma/{\rm d}t$ ratio at the bump to that at the dip, $R$, has been determined with similar MC studies.

    \begin{table*}\small\color{black}
        \begin{center}
            \caption{The differential cross-section ${\rm d}\sigma/{\rm d}t$ of DS1 at 2.76 TeV, measured at 13~$\sigma_{\rm beam}$ distance.}
            \begin{tabular}{ | c  c  c | c  c  c|}
		\hline
			$|t|_{\rm low}$		& $|t|_{\rm high}$		&	$|t|_{\rm repr.}$		&	${\rm d}\sigma/{\rm d}t$	& Statistical uncertainty 	&	Systematic uncertainty \\ 
						& [GeV$^{2}$]			&					&					&	[mb GeV$^{-2}$]		& 				\\ \hline 
	0.3625	&	0.3825	&	0.37190	&	0.6565	&	0.0277	&	0.0331\\
	0.3825	&	0.4025	&	0.39188	&	0.4536	&	0.0183	&	0.0238\\
	0.4025	&	0.4225	&	0.41185	&	0.3133	&	0.0140	&	0.0168\\
	0.4225	&	0.4425	&	0.43184	&	0.2075	&	0.0091	&	0.0116\\
	0.4425	&	0.4625	&	0.45185	&	0.1270	&	0.0062	&	0.0073\\
	0.4625	&	0.4825	&	0.47188	&	0.0952	&	0.0056	&	0.0056\\
	0.4825	&	0.5025	&	0.49189	&	0.0718	&	0.0054	&	0.0042\\
	0.5025	&	0.5225	&	0.51186	&	0.0413	&	0.0044	&	0.0024\\
	0.5225	&	0.5425	&	0.53181	&	0.0319	&	0.0041	&	0.0020\\
	0.5425	&	0.5625	&	0.55180	&	0.0209	&	0.0033	&	0.0013\\
	0.5625	&	0.5825	&	0.57190	&	0.0159	&	0.0029	&	0.0009\\
	0.5825	&	0.6025	&	0.59213	&	0.0098	&	0.0021	&	0.0004\\
	0.6025	&	0.6225	&	0.61239	&	0.0085	&	0.0019	&	0.0004\\
	0.6225	&	0.6425	&	0.63295	&	0.0092	&	0.0017	&	0.0004\\
	0.6425	&	0.6625	&	0.65298	&	0.0131	&	0.0031	&	0.0008\\
	0.6625	&	0.6825	&	0.67266	&	0.0171	&	0.0047	&	0.0008\\
	0.6825	&	0.7225	&	0.70214	&	0.0148	&	0.0027	&	0.0008\\
	0.7225	&	0.7625	&	0.74119	&	0.0122	&	0.0030	&	0.0008\\ \hline
            \end{tabular}
        \label{DS1_data}
        \end{center}
    \end{table*}

    \begin{table*}\small\color{black}
        \begin{center}
            \caption{The differential cross-section ${\rm d}\sigma/{\rm d}t$ of DS2 at $\sqrt{s}=$2.76 TeV, measured at 4.3~$\sigma_{\rm beam}$ distance.}
            \begin{tabular}{ | c  c  c | c  c  c|}
		\hline
			$|t|_{\rm low}$		& $|t|_{\rm high}$		&	$|t|_{\rm repr.}$		&	${\rm d}\sigma/{\rm d}t$	& Statistical uncertainty 	&	Systematic uncertainty \\ 
						& [GeV$^{2}$]			&					&					&	[mb GeV$^{-2}$]		& 				\\ \hline 
0.0700	&	0.0750	&	0.07246	&	113.88	&	2.960	&	1.455	\\
0.0750	&	0.0800	&	0.07746	&	101.14	&	2.181	&	1.319	\\
0.0800	&	0.0850	&	0.08246	&	88.78	&	1.752	&	1.181	\\
0.0850	&	0.0900	&	0.08746	&	83.25	&	1.600	&	1.130	\\
0.0900	&	0.0950	&	0.09246	&	79.25	&	1.542	&	1.097	\\
0.0950	&	0.1000	&	0.09746	&	69.81	&	1.326	&	0.986	\\
0.1000	&	0.1050	&	0.10246	&	64.27	&	1.260	&	0.926	\\
0.1050	&	0.1100	&	0.10746	&	56.75	&	1.133	&	0.833	\\
0.1100	&	0.1150	&	0.11246	&	54.96	&	1.085	&	0.822	\\
0.1150	&	0.1200	&	0.11746	&	49.43	&	1.002	&	0.754	\\
0.1200	&	0.1250	&	0.12246	&	46.41	&	1.135	&	0.721	\\
0.1250	&	0.1300	&	0.12746	&	41.59	&	0.904	&	0.658	\\
0.1300	&	0.1350	&	0.13246	&	39.28	&	0.853	&	0.633	\\
0.1350	&	0.1400	&	0.13746	&	34.87	&	0.782	&	0.572	\\
0.1400	&	0.1450	&	0.14246	&	33.42	&	0.775	&	0.558	\\
0.1450	&	0.1500	&	0.14746	&	30.92	&	0.746	&	0.526	\\
0.1500	&	0.1550	&	0.15246	&	26.78	&	0.666	&	0.463	\\
0.1550	&	0.1600	&	0.15746	&	25.73	&	0.645	&	0.453	\\
0.1600	&	0.1650	&	0.16246	&	23.32	&	0.602	&	0.417	\\
0.1650	&	0.1700	&	0.16746	&	20.63	&	0.573	&	0.376	\\
0.1700	&	0.1750	&	0.17246	&	19.84	&	0.564	&	0.367	\\
0.1750	&	0.1800	&	0.17746	&	18.14	&	0.520	&	0.341	\\
0.1800	&	0.1850	&	0.18246	&	16.68	&	0.493	&	0.319	\\
0.1850	&	0.1900	&	0.18746	&	15.61	&	0.490	&	0.303	\\
0.1900	&	0.1950	&	0.19246	&	13.99	&	0.459	&	0.276	\\
0.1950	&	0.2000	&	0.19746	&	12.55	&	0.426	&	0.251	\\
0.2000	&	0.2050	&	0.20246	&	11.21	&	0.423	&	0.228	\\
0.2050	&	0.2100	&	0.20746	&	11.56	&	0.542	&	0.239	\\
0.2100	&	0.2150	&	0.21246	&	9.59	&	0.368	&	0.201	\\
0.2150	&	0.2200	&	0.21746	&	8.95	&	0.358	&	0.190	\\
0.2200	&	0.2250	&	0.22246	&	8.34	&	0.340	&	0.180	\\
0.2250	&	0.2300	&	0.22746	&	7.47	&	0.307	&	0.164	\\
0.2300	&	0.2350	&	0.23246	&	7.35	&	0.322	&	0.163	\\
0.2350	&	0.2400	&	0.23746	&	6.72	&	0.307	&	0.151	\\
0.2400	&	0.2450	&	0.24246	&	6.28	&	0.291	&	0.143	\\
0.2450	&	0.2500	&	0.24746	&	5.58	&	0.269	&	0.129	\\
0.2500	&	0.2750	&	0.26161	&	4.08	&	0.103	&	0.098	\\
0.2750	&	0.3000	&	0.28661	&	2.87	&	0.089	&	0.074	\\
0.3000	&	0.3250	&	0.31161	&	1.66	&	0.064	&	0.045	\\
0.3250	&	0.3500	&	0.33661	&	1.15	&	0.056	&	0.033	\\
0.3500	&	0.3750	&	0.36161	&	0.82	&	0.044	&	0.025	\\
0.3750	&	0.4000	&	0.38661	&	0.47	&	0.036	&	0.015	\\
0.4000	&	0.4250	&	0.41161	&	0.34	&	0.045	&	0.012	\\
0.4250	&	0.4500	&	0.43661	&	0.19	&	0.024	&	0.007	\\
0.4500	&	0.4750	&	0.46161	&	0.13	&	0.018	&	0.005	\\ \hline
            \end{tabular}
        \label{DS2_data}
        \end{center}
    \end{table*}

\section{Discussion of the results}
\label{discussion_of_physics_results}

The TOTEM experiment at CERN LHC has observed the presence of a diffractive minimum at $\sqrt{s}=2.76$~TeV in elastic pp scattering with high significance. The importance of this observation is that the new data measured at $\sqrt{s}=2.76$~TeV
are rather close in energy to the D0 results of p$\bar{\rm p}$ measured at $\sqrt{s}=1.96$~TeV. Apparently, the dynamics of pp and p$\rm\bar{p}$ change on the scale of $\ln{\sqrt{s}}$, to be studied in detail in
a forthcoming TOTEM and D0 collaboration analysis. 

The measured ratio of the pp differential cross-section at the bump and dip is $R=1.7 \pm 0.2$, see Fig.~\ref{differential_cross_section_normalization}. The pp data also shows a steepening of the differential
cross-section and a change in the nuclear slope $B(t)$ starting at $|t|\approx0.4$~GeV$^{2}$ (Fig.~\ref{differential_cross_section_with_D0_fits}). Both features are 
absent in the p$\bar{\rm p}$ measured at $\sqrt{s}=1.96$~TeV of the D0 experiment, where a kink structure without a minimum and a subsequent maximum can be observed with $R_{\rm p\bar{p}}=1.0\pm 0.1$. This value $R_{\rm p\bar{p}}$ and its uncertainty
can be obtained by fitting the published D0 data in the $t$-range of the plateau (nearly constant ${\rm d}\sigma/{\rm d}t$), including and after the kink~\cite{Abazov:2012qb}. Therefore, the incompatibility on the $R$ parameter between the p$\bar{\rm p}$ data at $\sqrt{s}=1.96$~TeV
and the pp data at 2.76~TeV is approximately 3$\sigma$.

At higher LHC energies of 7~TeV and 13~TeV, the dip has been observed already earlier by TOTEM. It is evidently a permanent structure in pp elastic scattering 
at LHC energies.

As far as we know, there are no models which are able to describe the pp TOTEM data and the $\rm p\bar{p}$ D0 data (total cross-section,
$\rho$, dip-region) without the effects of the Odderon~\cite{Cudell:2002xe,Khoze:2017swe,Khoze:2013jsa}. On the contrary, theoretical models including the effects of the Odderon have predicted the observed effects and are able to describe both the pp TOTEM data and the $\rm p\bar{p}$ D0 data (TeV scale)~\cite{Lukaszuk:1973nt,Martynov:2017zjz,Gauron:1992zc}.

Therefore, unless something unknown happens between $\sqrt{s}=$2.76~TeV and 1.96~TeV, the significant difference between the pp and $\rm p\bar{p}$ differential cross-section provides evidence for a colourless
3-gluon bound state exchange in the $t$-channel of the proton-proton elastic scattering.

The observed difference between pp and $\rm p\bar{p}$ is the most classic definition of evidence for the Odderon since the last day of run at the ISR (more than 40 years ago)~\cite{Breakstone:1985pe,Breakstone:1984te,Amaldi:1979kd}.
While at lower energies the diffractive dip contributions may naturally come from secondary Reggeons, their contribution is generally considered negligible, less than 1~\%, at LHC energies due to their Regge trajectory intercept
lower than unity~\cite{Jenkovszky:2017efs}.

A variety of odd-signature exchanges relevant at high energies have been discussed in literature, within different frameworks and under different names, see e.g. the review~\cite{Ewerz:2005rg}.
The ``Odderon'' was introduced within
the axiomatic theory~\cite{Gauron:1992zc} as an amplitude contribution responsible for the difference between pp and $\rm p\bar{p}$
differential cross-section in the dip region. Crossing-odd trajectories were also studied
within the framework of Regge theory as a counterpart of the crossing-even Pomeron.  It has also been shown
that such object must exist in QCD, as a colourless bound state of three gluons with quantum numbers $J^{PC}=1^{--}$ (see e.g.~\cite{Bartels:1999yt}).
The binding strength among the 3 gluons is greater than the strength of their interaction with other particles. There is also
evidence for such a state in QCD lattice calculations, known under the name  “vector glueball” (see e.g.~\cite{Morningstar:1999rf}).
However, in lattice QCD, Odderon effects were calculated only without dynamical quarks so far.
Such a state, on one hand, can be exchanged in the $t$-channel and contribute, e.g., to the elastic scattering amplitude.
On the other hand it can be created in the $s$-channel and thus be observed in spectroscopic studies as well.

There are multiple ways how an odd-signature exchange component may manifest itself in observable data.  Focussing on elastic scattering at
the LHC (unpolarised beams), there are 3 regions often argued to be sensitive.  In general, the effects of an
odd-signature exchange (bound state of 3 or odd number of gluons) are expected to be much smaller than those of even-signature exchanges (bound state of 2 or even number of gluons).
Consequently, the sensitive regions are those where the contributions from 2-bound-gluon exchanges cancel or are small.  At very
low-$|t|$ the 2-bound-gluon amplitude is expected to be almost purely imaginary, while a 3-gluon exchange would make contributions
to the real part and therefore $\rho$, the ratio of the real to imaginary part of the nuclear scattering amplitude, is a very sensitive parameter. The $\sigma_{\rm tot}$ and $\rho$ parameter measurement of the TOTEM experiment at $\sqrt{s}=13$~TeV already provided the first indication for
the existence of a colourless 3-gluon bound state~\cite{Antchev:2017dia,Antchev:2298154}.

Sometimes the high-$|t|$ region is also argued to be sensitive to 3-gluon exchanges since the contribution from 2-gluon exchanges is rapidly
decreasing.  However, preliminary high-$|t|$ TOTEM data at 13 TeV indicate that this region is dominated either by a perturbative-QCD amplitude, see e.g.~\cite{Donnachie:1979yu},
and high energy predictions~\cite{Brodsky:1973kr}.

The third opportunity is the comparison of the dip range, exploited in this analysis. The dip is often described as the $t$-range where the imaginary part of the amplitude is crossing zero, thus ceding
the dominance to the real part to which a 3-gluon exchange may contribute.  In agreement with such predictions, the observed dips
in $\rm p\bar{p}$ scattering are shallower than those in pp. There are data at $\sqrt{s}=$53 GeV showing a significant difference between the pp and $\rm p\bar{p}$
dip~\cite{Amaldi:1979kd}.  The interpretation of this difference is, however, complicated due to a possible non-negligible contribution from secondary Reggeons.
These are not expected to give sizeable effects at the Tevatron energies, which thus gives weight to the D0 observation of a very shallow
dip in $\rm p\bar{p}$ elastic scattering~\cite{Abazov:2012qb} compared to the very pronounced dip measured by TOTEM at 2.76~TeV, 7~TeV and 13~TeV~\cite{Antchev:2011vs}.

\section{Summary}

The proton-proton elastic differential cross section ${\rm d}\sigma/{\rm d}t$ has been measured by the TOTEM experiment at $\sqrt{s}=2.76$~TeV LHC energy with $\beta^{*}=11$~m beam optics. 
The differential cross-section can be described with an exponential in the range $0.36<|t|<0.54$~GeV$^{2}$, followed by a significant diffractive minimum at $|t_{\rm dip}|=(0.61\pm0.03)$~GeV$^{2}$.
The ratio of the ${\rm d}\sigma/{\rm d}t$ between the bump (the local maximum subsequent to the dip) and dip is $R=1.7\pm0.2$. This value $R$ is significantly different from $R_{\rm p\bar{p}}=1.0\pm 0.1$, obtained from the $\rm p\bar{p}$ measurement of
the D0 experiment at $\sqrt s = 1.96\un{TeV}$. 

Neglecting the small energy difference in $\sqrt{s}$ between the measurements of the TOTEM and D0 collaborations~\cite{Abazov:2012qb}, the results provide evidence for a colourless 3-gluon bound state exchange in the $t$-channel of the proton-proton elastic scattering.
The presented observables $R$ and the nuclear slope before the dip $B_{\rm pp}$ ($B_{\rm p\bar{p}}$) are both $\sqrt{s}$ dependent, and this dependence has to be studied in detail in order to quantify the exact significance of the observation.
This will be the subject of a forthcoming joint publication by the TOTEM and D0 experiments.

\section*{Acknowledgments}

We are grateful to the beam optics development team for providing the beams and operating the instrumentation, to the machine coordinators
and the LPC coordinators for scheduling the dedicated fills.

This work was supported by the institutions listed on the
front page and partially also by NSF (US), the Magnus Ehrnrooth Foundation (Finland), the Waldemar von
Frenckell Foundation (Finland), the Academy of Finland,
the Finnish Academy of Science and Letters (The Vilho 
Yrj\"o and Kalle V\"ais\"al\"a Fund), the Circles of Knowledge Club (Hungary) and the OTKA NK 101438 and the EFOP-3.6.1-16-2016-00001 grants
(Hungary). Individuals have received support from Nylands nation vid Helsingfors universitet (Finland),
MSMT CR (the Czech Republic), the J\'{a}nos Bolyai Research Scholarship of
the Hungarian Academy of Sciences, the NKP-17-4 New National Excellence Program of the
Hungarian Ministry of Human Capacities and the Polish Ministry of Science and Higher Education
Grant no. MNiSW DIR/WK/2017/07-01.

\clearpage
\bibliographystyle{utphys}
\bibliography{mybib}

\end{document}